\documentclass[aps]{revtex4}
\usepackage{amsfonts}
\usepackage{amsmath}
\usepackage{amssymb}
\usepackage{graphicx}

\begin{document}

\title{Sub-Rayleigh resolution of interference pattern with independent
laser beams}
\author{Jun Xiong, Feng Huang, De-Zhong Cao, Hong-Guo Li, Xu-Juan Sun,
and Kaige Wang$\footnote{Corresponding author:
wangkg@bnu.edu.cn}$}

\affiliation{Department of Physics, Applied Optics Beijing Area
Major Laboratory, Beijing Normal University, Beijing 100875,
China}

\begin{abstract}
We propose a new scheme to achieve sub-Rayleigh resolution of interference
pattern with independent laser beams. We perform an experimental observation
of a double-slit interference with two orthogonally polarized laser beams.
The resolution of the interference pattern measured by a two-photon
detection is doubled provided the two beams illuminate the double-slit with
certain incident angles. The scheme is simple and can be in favor of both
high intensity and perfect visibility.
\end{abstract}

\maketitle

The resolution of optical pattern is restricted by light
diffraction and the limit is described by the Rayleigh criterion.
Many
efforts have been devoted to surpassing this limit\cite{ya}$^{-}$\cite{xj}%
for purpose of improving photolithography technology and micro-structure
analysis. Among them quantum schemes using entangled photons can perform
sub-wavelength interference, resulting in essence the resolution enhancement.%
\cite{fon}$^{-}$\cite{shih} However, the practical implementation of this
kind of lithography is challenging because of the very low efficiency. The
further studies showed that sub-wavelength interference can exist in the
macroscopic realm such as a pair of entangled beams generated in a high gain
spontaneous parametric down-conversion (SPDC)\cite{nag,wang} and a thermal
light source\cite{wang1,xj}, where the interference patterns can be much
intensive. But as an expense, the fringe visibility becomes lower.
Naturally, an optimum scheme of sub-Rayleigh resolution should comprise both
high intensity and perfect visibility of interfered pattern.

In this paper we propose a scheme of sub-Rayleigh resolution which may
satisfy the requirements above. The scheme is classical, using two
independent laser beams illuminating a double-slit with appropriate incident
angles. The interference pattern is recorded by a two-photon exposure system
which simultaneously absorbs one photon from each beam. The pattern
resolution is doubled with perfect visibility while the intensity is
proportional to that of laser beams.

Let the interference pattern be described by $I(x,\phi )\propto 1+\cos
(kdx/z-\phi )$, where $k=2\pi /\lambda $ is the wavevector of the beam, $d$
the slit space, and $z$ the distance between the double-slit and the
detection plane. $\phi =$ $kd\tan \theta $ is a phase shift introduced by an
incident angle $\theta $ of the beam with respect to the double-slit. Here
we omit the envelope of the pattern. If the two orthogonally polarized beams
with the same wavelength illuminate the double-slit, the intensity product
of the two patterns is given by
\begin{eqnarray}
I_{1}(x_{1,}\phi _{1})I_{2}(x_{2,}\phi _{2}) &\propto &1+\cos
(kdx_{1}/z-\phi _{1})+\cos (kdx_{2}/z-\phi _{2})  \nonumber \\
&&+(1/2)\cos [kd(x_{1}+x_{2})/z-(\phi _{1}+\phi _{2})]  \nonumber \\
&&+(1/2)\cos [kd(x_{1}-x_{2})/z-(\phi _{1}-\phi _{2})].  \label{1}
\end{eqnarray}%
The second and third terms in Eq. (\ref{1}) show the normal resolution of
the pattern whereas the third or fourth term may exhibit a superresolution
pattern. For the source of entangled photons, the quantum interference
eliminates the low resolution term. As for the classical schemes, such as a
thermal correlation source\cite{wang1,xj} and the source consisting of beams
with different frequencies\cite{ya}, the low resolution part is averaged
causing a background. In the present model, however, the low resolution
terms can be eliminated by setting the appropriate phase shifts. For the
cases of $x_{1}=x_{2}=x$ and $x_{1}=-x_{2}=x$, we obtain
\begin{eqnarray}
I_{1}(x_{,}\phi _{1})I_{2}(\pm x_{,}\phi _{2}) &\propto &(1/2)\{1+\cos
[2kdx/z-(\phi _{1}\pm \phi _{2})]\},\qquad  \nonumber \\
\text{when }\phi _{1}\mp \phi _{2} &=&(2n+1)\pi.  \label{2}
\end{eqnarray}%
This shows a superresolution pattern with perfect visibility. For
simplicity, let $\phi _{1}=-\phi _{2}=\phi $ for the case $x_{1}=x_{2}=x$
and $\phi _{1}=\phi _{2}=\phi $ for the case $x_{1}=-x_{2}=x$, Eq. (\ref{2})
is written as
\begin{equation}
I_{1}(x_{,}\phi )I_{2}(\pm x_{,}\mp \phi )\propto (1/2)[1+\cos
(2kdx/z)],\qquad \text{when }\phi =(2n+1)\pi /2.  \label{3}
\end{equation}%
By exchanging the detection ways, we obtain
\begin{equation}
I_{1}(x_{,}\phi )I_{2}(\mp x_{,}\mp \phi )\propto \lbrack 1+\cos (kdx/z-\pi
/2)]^{2},  \label{4}
\end{equation}%
which shows a normal resolution pattern.

The experimental setup is very simple, shown in Fig.1. Two orthogonally
polarized beams illuminate a double-silt with slit width $b=75$ $\mu $m and
slit distance $d=200$ $\mu $m. The diffracted radiation is split into two
beams with a 50/50 polarizing beam splitter(PBS). The transmitted and
reflected beams are then detected by small-area (diameter 0.4mm)
Si-photodetectors D$_{1}$ and D$_{2}$, which are mounted on translation
stages distant from the double-slit $z=770$ mm. We consider two kinds of
observation, $x_{1}=x_{2}=x$ and $x_{1}=-x_{2}=x,$ which are mimic to the
schemes of sub-wavelength interference for a two-photon entangled source\cite%
{shih} and a thermal light source\cite{wang1,xj}, respectively. The signals
of the two detectors are recorded on a digital oscilloscope (Tektronics
3012B). The experimental results are plotted in Figs. 2-5. In Figs. 2 and 3,
the two orthogonally polarized beams illuminate the double-slit with angles $%
\theta $ and $-\theta $, and in Figs. 4 and 5, the two beams with the same
angle $\theta $. Figure 2a (4a) shows individually the two interference
patterns, $I_{1}(x,\phi )$ and $I_{2}(x,-\phi )$ ($I_{2}(-x,\phi )$). Their
intensity product $I_{1}(x,\phi )I_{2}(x,-\phi )$ ($I_{1}(x,\phi
)I_{2}(-x,\phi )$) plotted in Fig. 2b (4b) exhibits the sub-Rayleigh
resolution of the pattern. Obviously, the effect can be simply understood by
the relative shift of a half fringe interval between the two patterns.
However, $I_{1}(x,\phi )$ and $I_{2}(-x,-\phi )$ ($I_{2}(x,\phi )$) are
individually measured in Fig. 3a (5a), and their product $I_{1}(x,\phi
)I_{2}(-x,-\phi )$ ($I_{1}(x,\phi )I_{2}(x,\phi )$) is plotted in Fig. 3b
(5b). We can see that the pattern $I_{2}(x,-\phi )$ indicated by the
triangles in Fig. 2a is the mirror symmetry of the pattern $I_{2}(-x,-\phi )$
indicated by the triangles in Fig. 3a, so that the pattern $I_{2}(-x,-\phi )$
is synchronously modulated with the pattern $I_{1}(x,\phi )$ resulting in
the normal resolution for their product. The experimental results are in
good agreement with the theoretical analysis indicated by the solid and
dashed lines in the plots.

This method can be applied to the interference of a pair of coherent beams
in a plane\cite{ya,boto}, as shown in Fig. 6. Both horizontal and vertical
polarized beams perform individual interference patterns in the plane. When
a relative phase $\phi _{1}-\phi _{2}=\pi $ is introduced between the two
beams, the product of the two patterns exhibits the sub-Rayleigh resolution
feature. In principle, it can also realize an arbitrary superresolution of $%
\lambda /(2N)$, provided $N$ independent beams are arranged with a relative
phase shift $2\pi /N$ with each other. For the frequency-degenerate case,
combination of both polarized and spatial modes can gain the number of
independent beams.

In conclusion, we have demonstrated a new scheme for achieving resolution
enhancement of interference pattern. The scheme is simple without using
nonlinear source or quantum entangled source and is in favor of both high
intensity and perfect visibility.

This research was supported by the National Fundamental Research Program of
China Project No. 2001CB309310, and the National Natural Science Foundation
of China, Project No. 60278021 and No. 10074008.

Figure Captions

Fig. 1. Sketch of the experimental setup. The two orthogonally polarized
beams illuminate double-slit with certain angles (see the inset). The
outgoing beams are split by a polarization beam splitter (PBS) and then are
individually recorded in the detection plane.

Fig. 2. Two polarized beams illuminate the double-slit with certain incident
angles $\theta $ and $-\theta $ : (a) individual interference patterns for
the two polarized beams $I_1(x,\phi )$ and $I_2(x,-\phi )$ and (b) their
product $I_1(x,\phi )I_2(x,-\phi )$. The solid and dashed lines in Figs. 2-5
represent the numerical simulations.

Fig. 3. Same as in Fig. 2 but the two detectors placed at a pair symmetric
positions: (a) individual interference patterns $I_1(x,\phi )$ and $%
I_2(-x,-\phi )$ and (b) their product $I_1(x,\phi )I_2(-x,-\phi )$.

Fig. 4. Two polarized beams illuminate the double-slit with the same angle $%
\theta $ : (a) individual interference patterns for the two polarized beams $%
I_1(x,\phi )$ and $I_2(-x,\phi )$ and (b) their product $I_1(x,\phi
)I_2(-x,\phi )$.

Fig. 5. Same as in Fig. 4 but the two detectors placed at the same position:
(a) individual interference patterns $I_1(x,\phi )$ and $I_2(x,\phi )$ and
(b) their product $I_1(x,\phi )I_2(x,\phi )$.

Fig. 6. Sketch of two pairs (horizontal and vertical polarizations) of
coherent beams interfered in a plane. $\phi _{1}$ and $\phi _{2}$ are the
phase shifts in each pair of beams.

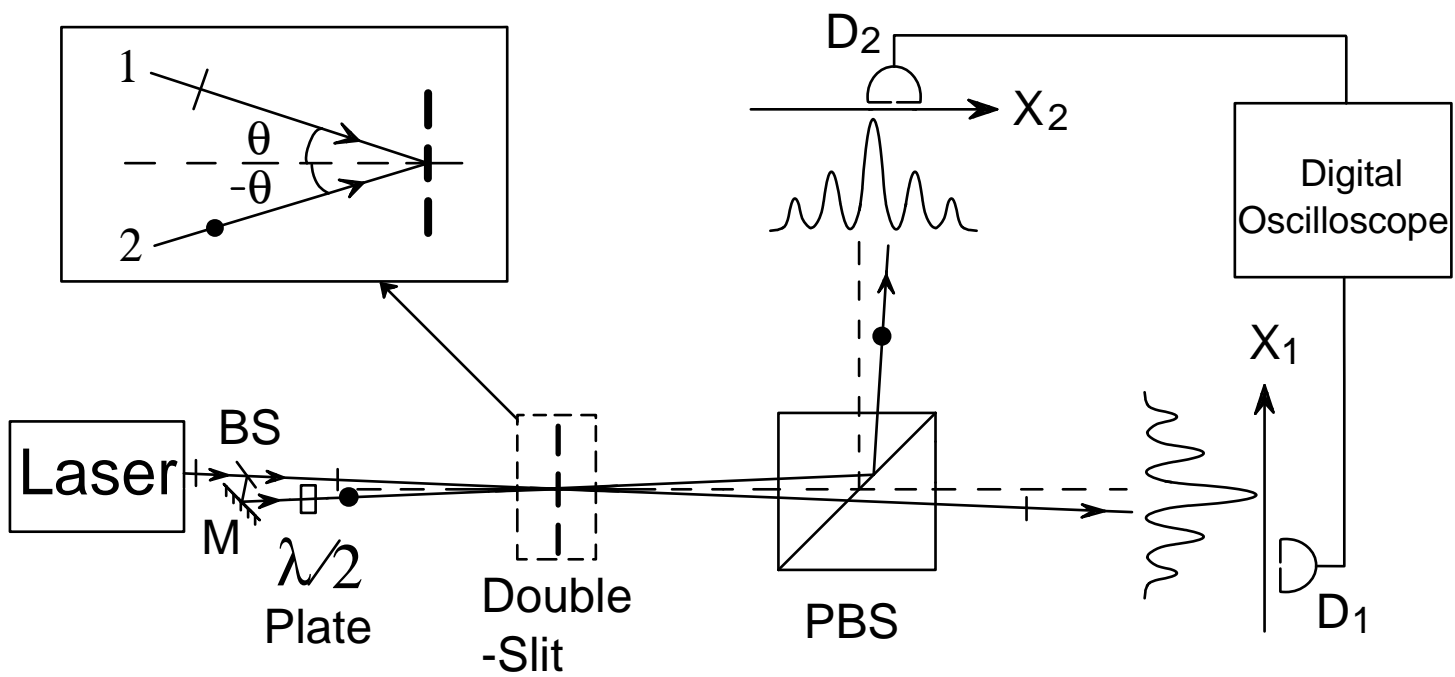
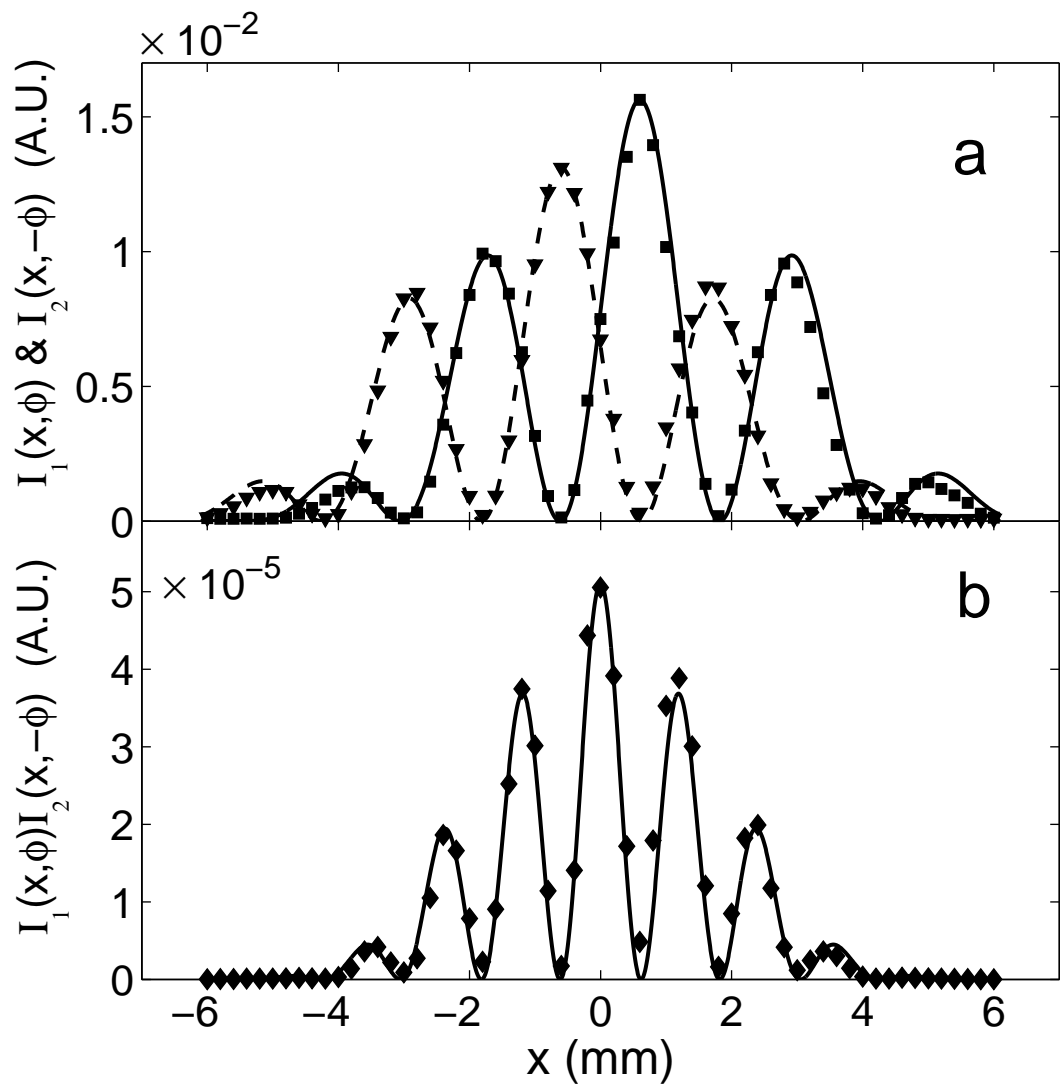
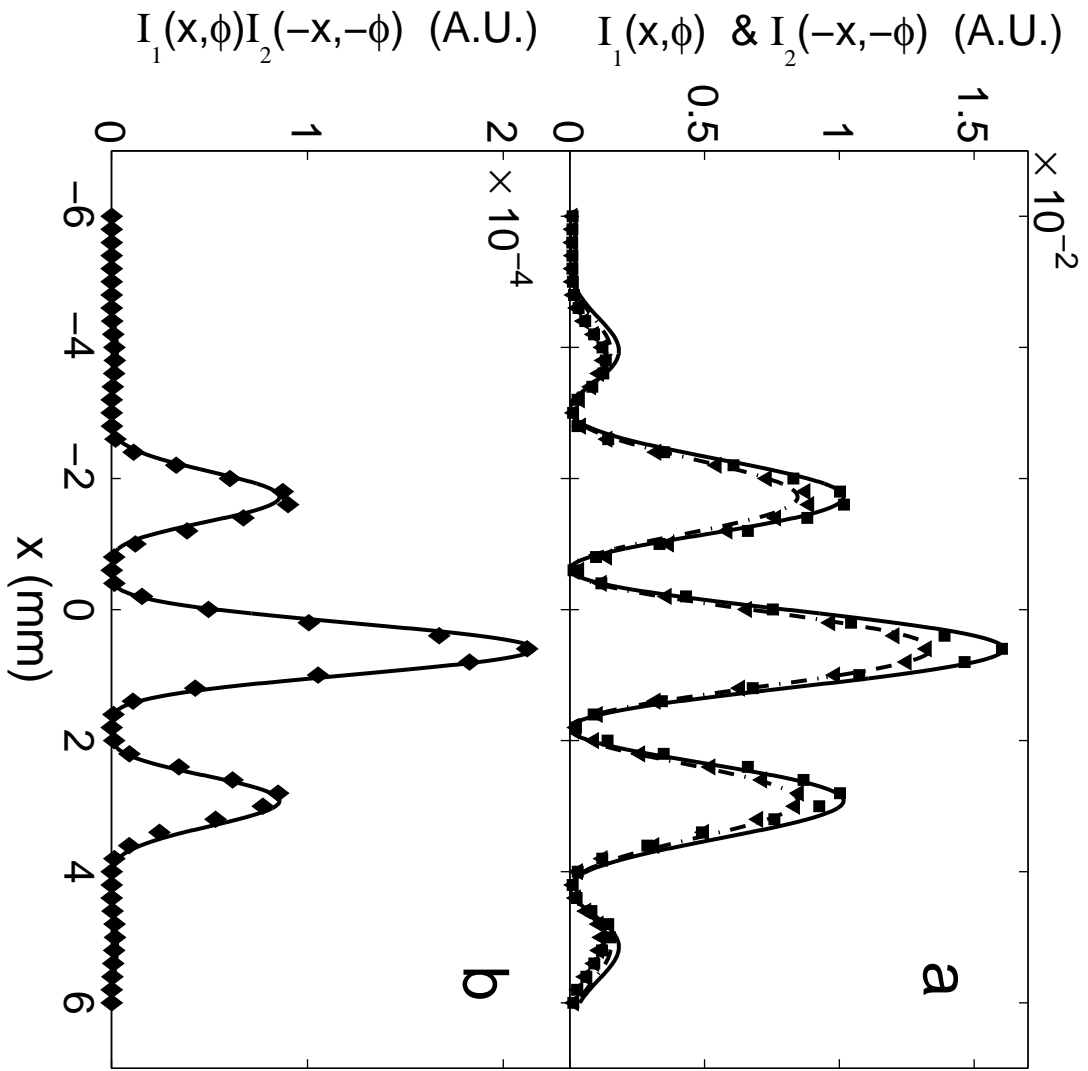
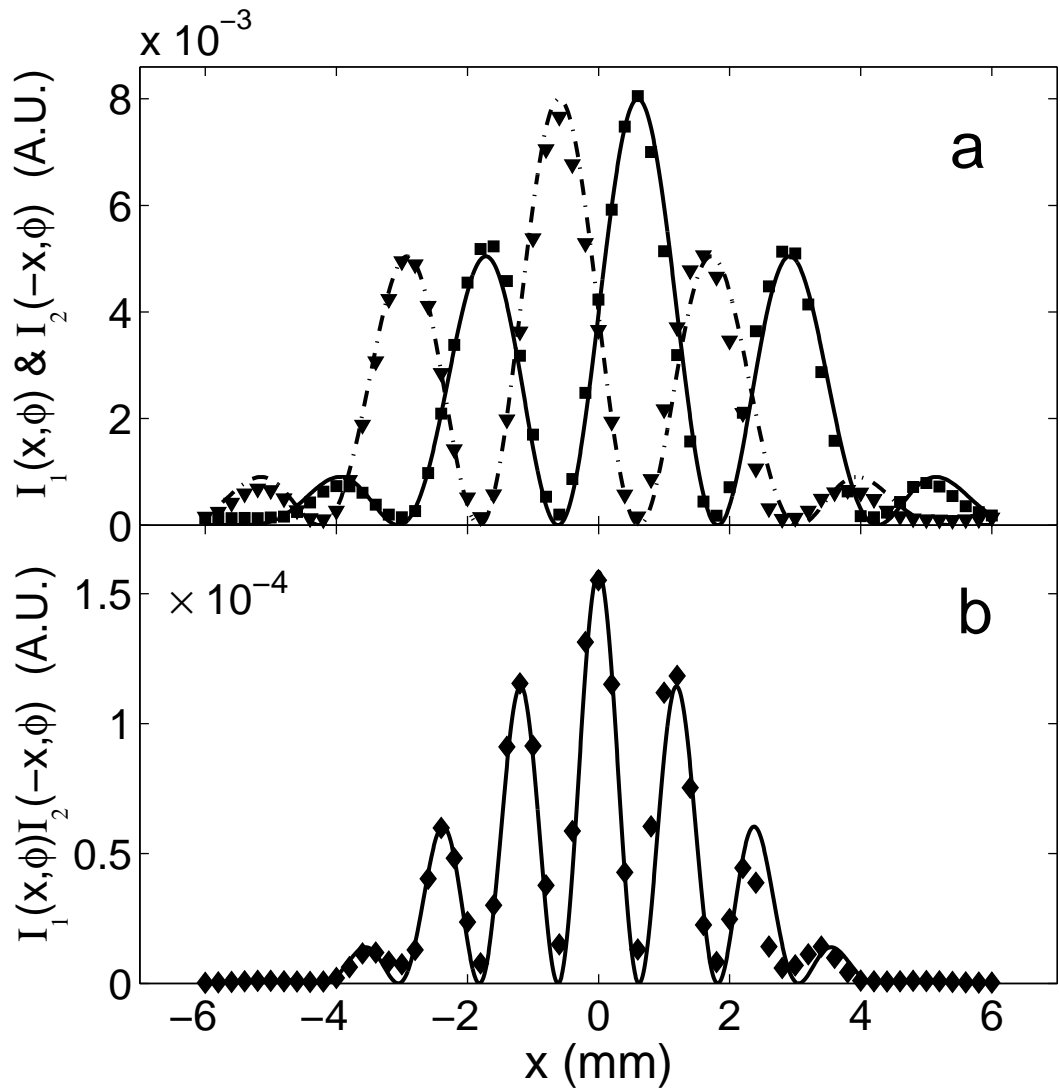
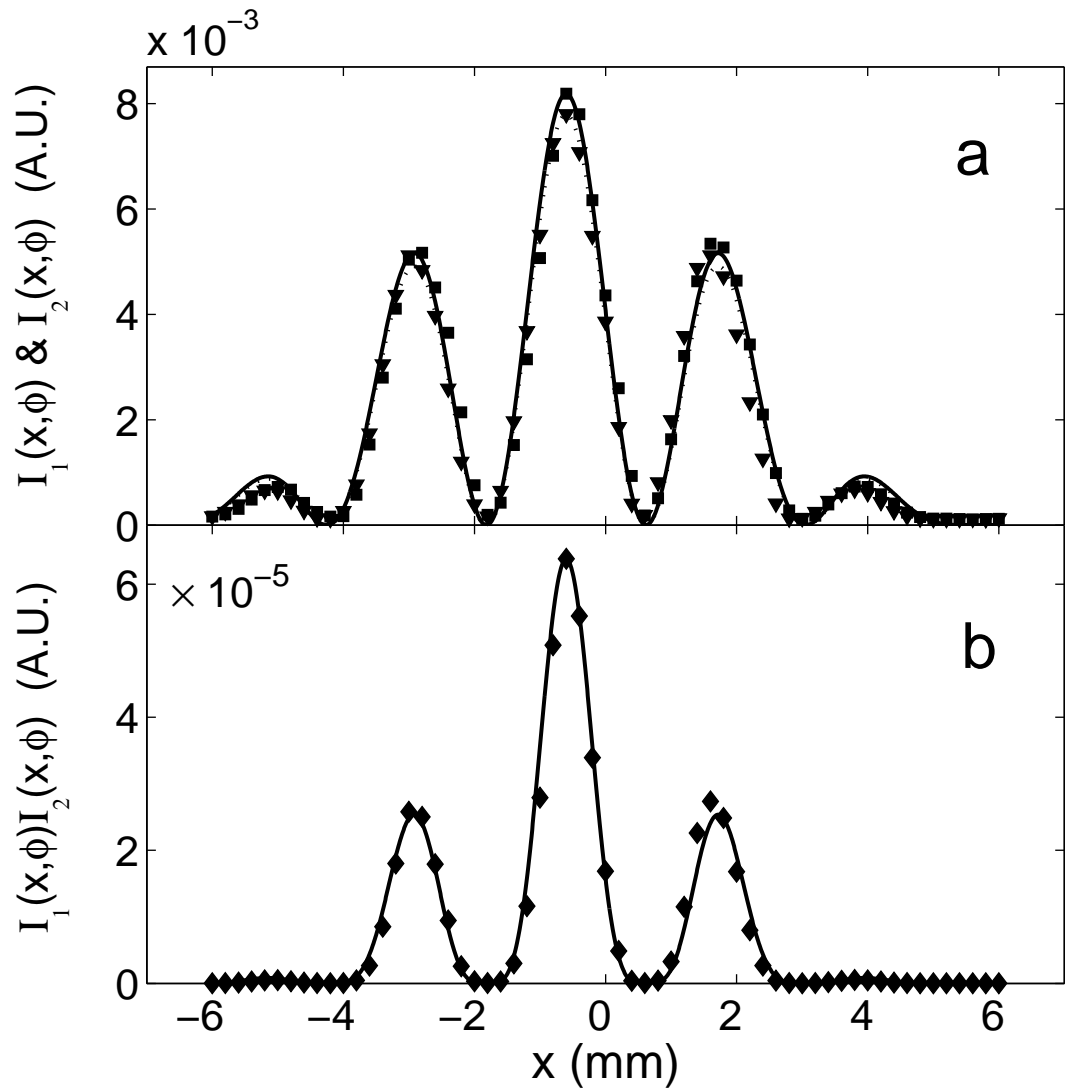
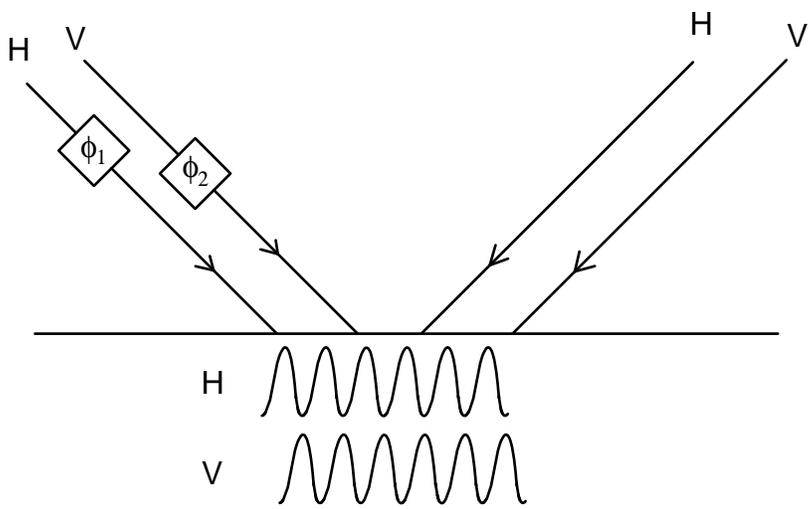
\end{document}